\documentclass[11pt]{book}
\pdfoutput=1
\usepackage[authoryear]{natbib}
\usepackage{graphicx}
\usepackage{setspace}

\title{\bf Low-frequency regime transitions and predictability of regimes
  in a barotropic model} 
\author{B. T. Nadiga and T.J. O'Kane} 
\date{\today}
\begin{document}
\frontmatter
\maketitle
\mainmatter
\onecolumn
\tableofcontents
\section{Abstract}
\index{Abstract} The predictability of flow is examined in a barotropic
vorticity model that admits low frequency regime transitions between
zonal and dipolar states. Low-frequency regime transitions in the
model were first studied by Bouchet and Simonnet (2009) and are
reminiscent of regime change phenomena in the weather and climate
systems wherein extreme and abrupt qualitative changes occur,
seemingly randomly, after long periods of apparent
stability. Mechanisms underlying regime transitions in the model
are not well understood yet. From the point of view of atmospheric and
oceanic dynamics, a novel aspect of the model is the lack of any
source of background gradient of potential-vorticity such as
topography or planetary gradient of rotation rate \citep[e.g., as
considered by][]{charney1979multiple}.

We consider perturbations that are embedded onto the system's chaotic
attractor under the full nonlinear dynamics as bred
vectors---nonlinear generalizations of the leading (backward) Lyapunov
vector.  We find that ensemble predictions that use bred vector
perturbations are more robust in terms of error-spread relationship
than those that use Lyapunov vector perturbations.  In particular,
when bred vector perturbations are used in conjunction with a simple
data assimilation scheme (nudging to truth) that estimates the current
state of the system, we find that at least some of the evolved
perturbations align to identify low-dimensional subspaces associated
with regions of large forecast error in the control (unperturbed,
data-assimilating) run; this happens less often in ensemble
predictions that use Lyapunov vector perturbations. Nevertheless, in
the inertial regime we consider, we find that (a) the system is more
predictable when it is in the zonal regime, and that (b) the horizon of
predictability is far too short compared to characteristic time scales
associated with processes that lead to regime transitions, thus
precluding the possibility of predicting such transitions.

\section{Introduction}

The phenomenon of blocking---large-scale patterns in the atmospheric
pressure field that are nearly stationary---in the extra-tropical
winter atmosphere \citep[e.g.,][]{charney1979multiple} is one of the
best studied cases of regime transitions that occur in the weather and
climate systems. Other such phenomena include bimodality of the
Kuroshio extension system \citep{qiu2000kuroshio} wherein beta and
topographic effects lead to two preferred meander patterns for the
Kuroshio current south of Japan and Dansgaard-Oeschger (D-O) events
\citep[e.g.,][]{dansgaard1989abrupt}---25 events during the last
glacial period that involved rapid warming followed by gradual cooling
with a recurrence time of about 1500 years.  

Clearly regime transitions can occur due to a variety of reasons. For
example, for atmospheric blocks, quasi-stationary Rossby wave trains
and synoptic-scale transient eddies are 
recognized as playing a role \citep[e.g., see ][and references
therein]{nakamura1997role}, whereas beta and topographic effects are
seen to be important in the bi-modality of the Kuroshio
current. (Causes of D-O events are less clear.) In the context of
atmospheric blocking, \cite{charney1979multiple} proposed a
multiple equilibria hypothesis whereby the atmosphere possesses
multiple stable steady states corresponding to observed multiple
weather regimes. The setting in this study and numerous others that
followed consisted of a barotropic model with topography. It is,
however, interesting to note that while topography (or other sources
of background gradients in potential vorticity) are not necessary for
the transient eddy mechanism mentioned above, most studies have indeed
included such sources of background gradients in potential
vorticity. We are not aware of studies of low-frequency regime
transitions that do not include such sources.  It is in this context
that the recent work of \cite{bouchet2009random} (referred to as BS09
for brevity) fills a gap: The model they consider has no background
gradient of potential vorticity (as due to a gradient of planetary
rotation rate or due to topography,) and therefore no (planetary or
topographic) Rossby waves.  Nevertheless, when the model is subjected
to weak stochastic forcing representative of inherently unpredictable,
or unresolved physics, irregular low-frequency zonal-dipolar regime
transitions arise.

The advantage of this model is that while on the one hand, its
behavior is reminiscent of regime change phenomena in the ocean and
climate systems wherein extreme and abrupt qualitative changes occur,
seemingly randomly, after very long periods of apparent stability, on
the other, it is simple enough that it is likely to yield to better
understanding of the dynamics underlying such phenomena. We note that
the dynamics of transitions in the model can be described in different
manners. For example, BS09 explain observed regime transitions in the
model in terms of phase transitions between dipolar structures and
unidirectional flows. (For details, the reader is referred to BS09.)
On the other hand, in an unforced setting,
\cite{loxley2013bistability} (hereafter referred to as LN13) consider
a simple theory for predicting quasi-steady states that combines a
maximum entropy principal with a nonlinear parameterization of the
vorticity-stream-function dependency. We briefly digress to discuss this
explanation of regime transitions since it instigates the possibility
that regime transitions can be predicted.

\begin{figure} 
\centering
\includegraphics[width=\textwidth]{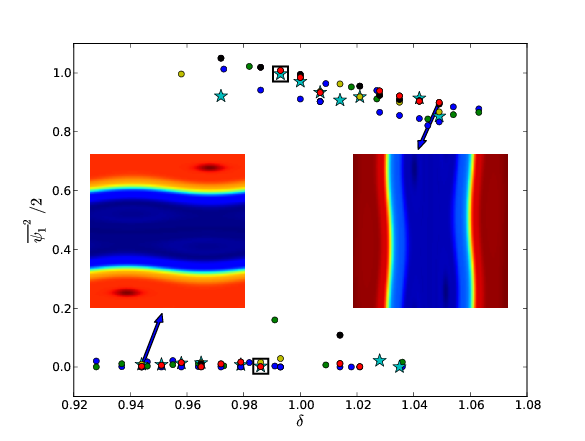}
\caption{The behavior of long-lived unidirectional flow states for
  different values of $\delta$ where $\delta$ is the square-root of
  the aspect ratio. A value of $\bar{\psi}_{1}^{2}/2\approx0$
  corresponds to unidirectional flow along the $x$-axis (left inset),
  while $\bar{\psi}_{1}^{2}/2\approx1$ corresponds to unidirectional
  flow along the $y$-axis (right inset). Note hysteresis and abrupt
  changes with respect to aspect ratio. Different symbols correspond
  to different sets of experiments where the initial conditions and/or
  small scale dissipation were varied. Adapted from LN13.}
\label{fig-bar}
\end{figure}

The theory of LN13 predicts that when the aspect ratio is varied,
unidirectional flows are bistable, exhibit hysteresis, and undergo
large abrupt changes in flow topology whereas dipolar structures
undergo continuous changes in flow topology.  These results are
summarized in Figs.~1 and 2 wherein values of $\bar\psi_1^2/2$ for
long-lived states at different values of aspect ratio are shown for a
large number of decaying turbulence simulations on a domain (0,
$2\pi\delta$) $\times$ (0, $2\pi/\delta$).  Here, $\bar\psi_1$ is the
amplitude of the eigenmode with the largest spatial scale in the
east-west or $x$ direction. In Fig.~1, the vorticity-stream-function
dependency is tanh-like and leads to unidirectional flows. In this
figure, a value of $\bar{\psi}_{1}^{2}/2\approx0$ corresponds to
unidirectional flow along the $x$-axis (Fig.~\ref{fig-bar}, left
inset), while $\bar{\psi}_{1}^{2}/2\approx1$ corresponds to
unidirectional flow along the $y$-axis (Fig.~\ref{fig-bar}, right
inset).  Different symbols correspond to different sets of experiments
where the initial conditions and/or small scale dissipation were
varied. Two branches of stable states are clearly seen in that figure:
the lower branch corresponds to unidirectional flow along the
$x$-axis; and the upper branch, to unidirectional flow along the
$y$-axis. These states coexist in a bistable region for some range of
$\delta$-values where $\delta$ is the square-root of the aspect ratio
(ratio of the size of the domain in the east-west or $x$ direction to
that in the north-south or $y$ direction).

\begin{figure}
\centering\includegraphics[width=\textwidth]{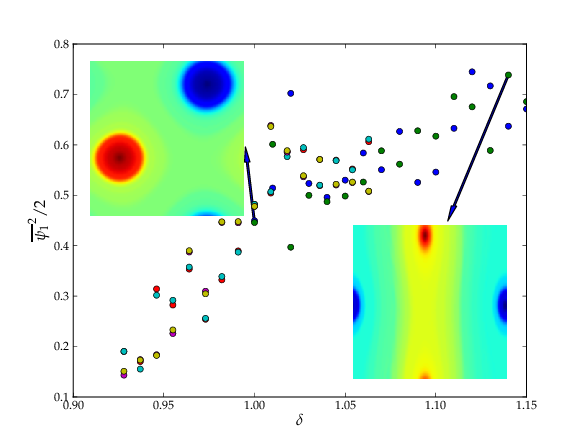}
\caption{The behavior of long-lived dipolar flow states for different
  values of $\delta$ where $\delta$ is the square-root of the aspect
  ratio. A value of $\bar{\psi}_{1}^{2}/2$ of 0.5 corresponds to a
  vortex pair with flow components along both the x- and y-axes (left
  inset), while deviations towards smaller or larger values indicate
  the tendency towards a more unidirectional flow (right inset). A
  more continuous change in flow is seen with respect to changes in
  aspect ratio. Adapted from LN13.}
\label{fig-dipole}
\end{figure}

In Fig.~2, the vorticity-stream-function dependency is sinh-like and
this leads to dipolar states. In this figure, a value of
$\bar{\psi}_{1}^{2}/2\approx0.5$ corresponds to a vortex pair with
flow components along both the $x$- and $y$-axes
(Fig.~\ref{fig-dipole}, Left Inset), while deviations towards smaller
or larger values of $\bar{\psi}_{1}^{2}/2$ indicate the tendency
towards a more unidirectional flow (e.g., Fig.~\ref{fig-dipole}, Right
Inset).  In contrast to Fig.~1, a single branch of stable states is
seen in Fig.~\ref{fig-dipole} as $\delta$ is varied. In the middle of
this branch is a single vortex pair in a square domain
(Fig.~\ref{fig-dipole}, left inset). As $\delta$ is increased, this
state is continuously ``squeezed" along the $y$-axis---eventually
yielding a large component of unidirectional flow along the $y$-axis
(Fig.~\ref{fig-dipole}, right inset). Similarly, decreasing $\delta$
eventually leads to a large component of unidirectional flow along the
$x$-axis. To a good approximation a continuous change in flow topology
from a vortex pair to a unidirectional flow takes place as $\delta$ is
changed: there is no sudden large change in flow topology as in
Fig.~\ref{fig-bar}.

To have large random changes between a unidirectional flow and a
dipolar structure requires both states to be quasi-steady
states. Except that they are quasi-steady states not at the same time,
but at nearby points in time.  Thus, the mechanism hypothesized by
LN13 to explain the random changes between a unidirectional flow and a
dipolar structure is as follows: Only one of the flow structures
(unidirectional flow or dipole) is a quasi-steady state over a short
timescale, and decaying turbulence causes relaxation towards that
state. Over a longer timescale, however, it is the cumulative effect
of the weak stochastic forcing that leads (randomly) to the other
state being the (now sole) preferred quasi-steady state and to which
it relaxes on the shorter timescale.  Indeed this mechanism leads to
the prediction that a regime transition is preceded by a change in
the kurtosis of the vorticity distribution. The question then arises
as to whether such behavior can be used to predict regime transitions
in the model. But, before considering that question, we characterize
differences in predictability of the zonal and dipolar regimes and
study the role of small scales in prediction and their role in
initiating regime transitions.

We now return from the digression. That abrupt changes that can occur
in the real weather and climate systems may be the result of small,
inherently unpredictable, or unresolved processes (e.g. see
\cite{frederiksen2008entropy}) is accounted for by the weak and
stochastic nature of the forcing in this model. In the parameter
regime we consider, the effect of the stochastic forcing when the
model is in one of the zonal or dipolar regimes itself is
minimal. However, the model displays no regime transitions in the
absence of stochastic forcing, and this is consistent with the
mechanism of LN13 described previously. Consequently, the occasional,
large (nonlinear) response of the system (regime shifts) to weak
stochastic forcing, and which response is unrelated to matching a
natural frequency of the system suggests a form of stochastic
resonance (\cite{benzi1982stochastic}, or e.g., see
\cite{williams2003spontaneous})---a constructive concurrence of
nonlinearity and noise---in setting up the regime transitions. In this
model, the underlying quasi-stable states essential to such a
mechanism are more dynamical than say the quasi-steady states of the
Kuroshio system. That is, whereas in the Kuroshio context, the
quasi-steady states are determined by specified topography, in the
present model the quasi-steady states are those anticipated by
statistical mechanical theories for the large scales of the system
(e.g., the Robert-Sommeria-Miller theory
\cite{chavanis1996classification}). Further, the difference in the
nature of the quasi-steady states is reflected (and captured) by
differences in the vorticity-streamfunction relationship and
differences in the kurtosis of the vorticity distribution itself.

\section{The barotropic vorticity model and its numerical
  discretization}
We consider the stochastically forced barotropic vorticity equation on
the f-plane:
\begin{equation}
\frac{D \omega}{D t} = \frac{\partial \omega}{ \partial t} + {\mathbf u}\cdot \nabla \omega = F + D
\label{eqn2}
\end{equation}
on a horizontal rectangular domain $2\pi\delta \times 2\pi/\delta $ with an
aspect ratio $\delta^2$.  Here, $\omega$ is the vertical
vorticity given in terms of streamfunction $\psi$ by
$\omega=\nabla^2\psi$, ${\mathbf u}$ is velocity given by ${\mathbf u}
= {\mathbf e_z} \times \nabla \psi$, $F$ is forcing and $D$ is
dissipation.  Dissipation consists of linear damping: $-\alpha\omega$,
where $\alpha$ is a frictional constant; and a small-scale dissipation
term that is a sink of the net-forward cascading enstrophy. The small
scale dissipation is implemented as a high pass filter of the
vorticity field with a characteristic wavenumber $k_d$ ($\approx 0.61
k_{max}$, where $k_{max}$ is the maximum wavenumber of the
simulation).  Forcing $F$ is scaled as $F=\sqrt{2\alpha}\tilde{F}$,
where $\tilde{F}$ is an isotropic stochastic forcing in a small band
of wavenumbers $2\leq k_f< 3$ drawn from independent unit variance
Gaussian distributions and which is temporally uncorrelated:
$\left<\tilde{F}_{\mathbf k}(t)\tilde{F}_{\mathbf
    k'}(t')\right>=\delta_{\mathbf k \mathbf k'}(t-t')$.  (A wide
variety of other spectral forcings considered produced similar
dipolar-zonal transitions.) If energy predominantly resides in the
large scales, then energy is mainly dissipated by linear damping. In
that case, the chosen scaling of forcing and linear damping terms
renders the long-time average of energy in a statistical stationary
state unity. Under these circumstances, the nondimensional time is
measured in terms of eddy turnover times where an eddy turnover time
is $L_{ref}/U_{ref}$ and given a dimensional setup $L_{ref}$ is such
that the nondimensional domain size is $2\pi \times 2\pi\delta$ and
$U_{ref}$ is such that the nondimensional energy is on average
unity. A fully-dealiased pseudo spectral spatial discretization, with
a 128x128 physical grid, is used in conjunction with an adaptive fourth order
Runge-Kutta time stepping scheme. The tendency of energy to cascade to
larger scales obviates the need for increased resolution; this was
verified using companion higher-resolution simulations.  In
order to observe regime transitions, the time evolution of higher
order moments of vorticity need to be adequately represented,
necessitating time steps that are O(100) times smaller than required
for stability.

In all simulations considered further, $\delta$ is fixed at
0.91. Simulations were considered for a number of values of $\alpha$
in the range $10^{-4}-10^{-3}$. However, results are presented for
the representative case of $\alpha=2\times 10^{-4}$. In particular
each simulation for that value of $\alpha$ was also repeated with
$\alpha=4\times 10^{-4}$ and the results were qualitatively unchanged.

\begin{figure}
\includegraphics[width=\textwidth]{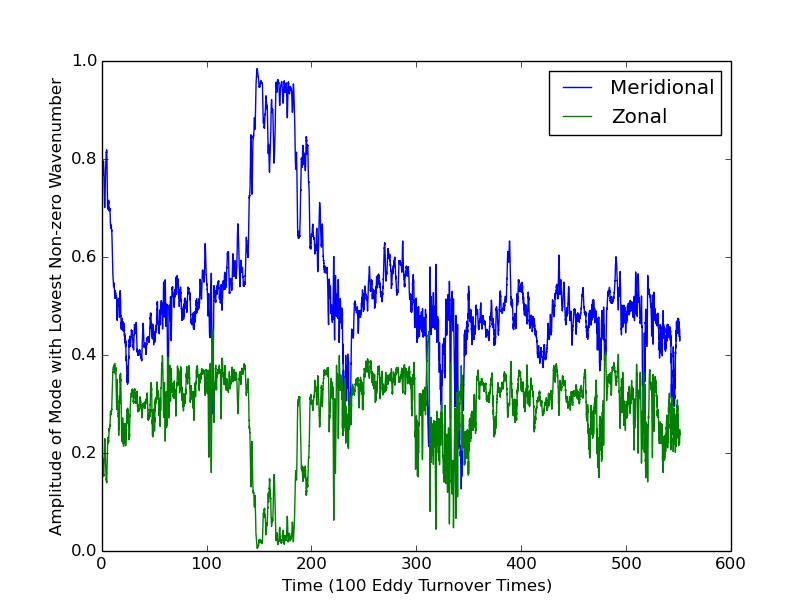}
\caption{Amplitude of zonal and meridional modes with the lowest
  non-zero wavenumber indicate transition from a dipolar state to a zonal
  state and back.}
\label{k1}
\end{figure}

\begin{figure}
\includegraphics[width=.49\textwidth]{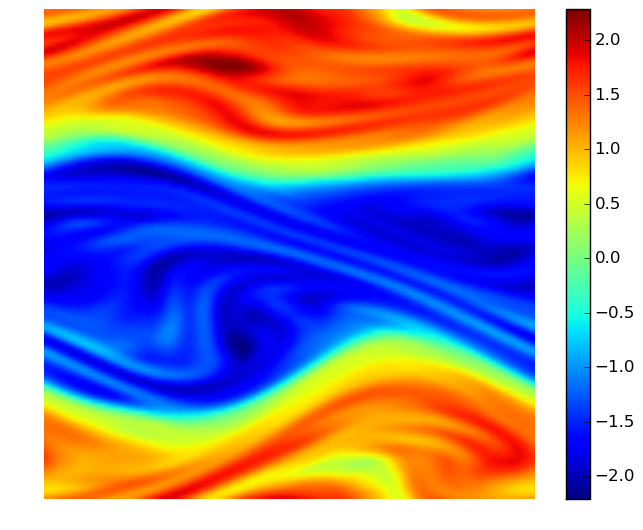}\hfill
\includegraphics[width=.49\textwidth]{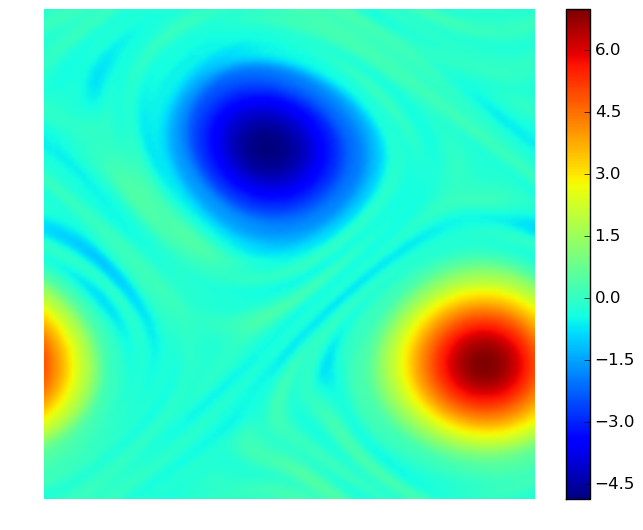}
\caption{Vorticity field when the system is in a zonal state (left, time=15000)
  and when the system is in a dipolar state (right, time=26000) in the reference run.}
\label{base-vort}
\end{figure}

\section{Low-frequency variability and regime transitions}
In the simulations considered, $\alpha$ variations in the range
$10^{-4}$ to $10^{-3}$ lead to dissipation timescales of the order of
$10^3$ eddy turnover times, and an energy input rate due to forcing
that is O($10^3$) times smaller than the nonlinear flux of energy in
the system. In this sense, forcing and dissipation are both weak
compared to nonlinearity in the dynamics of the model. Alternatively,
the weakness of the linear damping results in a damping scale for
energy that is larger than the size of the domain, allowing for energy
to condense on to the largest scales (e.g., see
\cite{smith1993bose}). Furthermore, in this regime, the model displays
behavior that is somewhat analogous to a subcritical pitchfork
bifurcation in the presence of noise (\cite{bouchet2009random}). That
is, under the influence of weak random forcing, the model switches
randomly and abruptly between zonal and dipolar states, as shown in
Figs.~\ref{k1} and \ref{base-vort}. 

Amplitude of modes with the lowest non-zero wavenumber in zonal and
meridional directions are plotted for a particular realization of the
weak stochastic forcing in Fig.~\ref{k1}. We will use this run as the
reference or truth run for further investigation of
predictability. The transition from a dipolar state to a zonal state
(around a time of 14000) and back (around a time of 20000) is evident
in that figure. The snapshot of vorticity at a time of 15000 (left
panel of Fig.~\ref{base-vort}) shows that the system has a larger
zonal flow component as compared to the state of the system at time 26000
(right panel of Fig.~\ref{base-vort}).  Consequently we will loosely
refer to states of the system that are similar to that at time 15000
as ``zonal'' and states similar to that at time 26000 as
``dipolar''. We also note that the kurtosis of the vorticity
distribution is negative and small in the zonal state whereas it is
positive and large in the dipolar state.

\section{Ensemble Perturbations}
Predictability is associated with the stability of the flow with
respect to perturbations (errors) and their associated growth.  For
infinitesimal initial perturbations, error growth may be close to
linear and the growing perturbations may be well described using a
tangent linear approximation to the full nonlinear evolution
equations. However, because of the possibility that the leading
Lyapunov exponent is related to small scale instabilities in a
multiscale system where predictability originates from the larger
scales (\cite{lorenz1969predictability, boffetta2002predictability}),
the leading Lyapunov exponent is often of limited relevance to
predictability. An associated common observation is that in
predictability studies using infinitesimal perturbations, linear
growth is seen over only an initial small fraction of the
predictability time. Therefore, what is more relevant to
predictability than the leading Lyapunov exponent is the evolution of
finite size perturbations under the full nonlinear equations.
Additionally, for a given norm, some perturbations with given spatial
structures may rapidly amplify whereas others will grow more slowly
and yet others decay.

\cite{pazo2010spatio} examined perturbation dynamics in extended
chaotic systems via the generically named ‘Lyapunov vectors’. They
demonstrated that one may, using the general method for the
calculation of Lyapunov vectors, generate initial perturbation vectors
that contain by construction different types of information about the
chaotic trajectory dependent on the initial perturbation magnitude
(infinitesimal or finite) and the evolution interval for the
calculation (finite or quasi-infinite; past, future or both).

Because of their ease of construction, knowledge of the past (flow
dependency) and dynamical balance we have chosen to characterize
space–time chaos in the model using bred vectors (BV). BVs are finite
perturbations generated (or ‘bred’) by evolving the perturbed system
$\omega'(t) = \omega(t)+\delta\omega(t)$ where $\omega(t)$ the control
trajectory under the full nonlinear governing equations. The
perturbations themselves are rescaled to a given size $\epsilon$
periodically at a time interval T as follows. The difference between
control and perturbed trajectories $\delta\omega(t+\delta t) =
\omega'(t+\delta t)-\omega(t+\delta t)$ is computed at times $\delta t
= nT$ for $n\in 1,\ldots,N\in\aleph$ whereupon the perturbation is
rescaled and the perturbed system redefined as $\omega'(t+\delta t) =
\omega(t+\delta t) + \epsilon\delta\omega(t+\delta
t)/||\delta\omega(t+\delta t)||$. The perturbation is now allowed to
evolve freely until the next rescaling is scheduled at time
$(n+1)T$. The bred vector corresponds to the (finite) perturbation
$\delta\omega(t)$ constructed at time t. We consider a range of values
for the perturbation amplitude and rescaling time. In our setting that
involves low frequency regime transitions, the rescaling time is
smaller than the average transition time by between two and three
orders of magnitude.

As control, we also consider backward Lyapunov vector perturbations
and compute them in the standard fashion
\citep{benettin1980lyapunov}. In this context, the amplitude of the
perturbation is chosen to be $10^{-5}$ times the enstrophy norm of the
fully equilibrated flow state to which perturbations are added and the
full nonlinear governing equations are used. The amplitude rescaling
time is chosen to be a characteristic eddy turnover time. As discussed
further later, when longer rescaling time periods are chosen, the
growth of perturbations leads to nonlinearities being significant. In
such cases, the behavior of the perturbations is qualitatively similar
to those of finite amplitude bred vector perturbations.

\begin{figure}
\centering
\includegraphics[width=\textwidth]{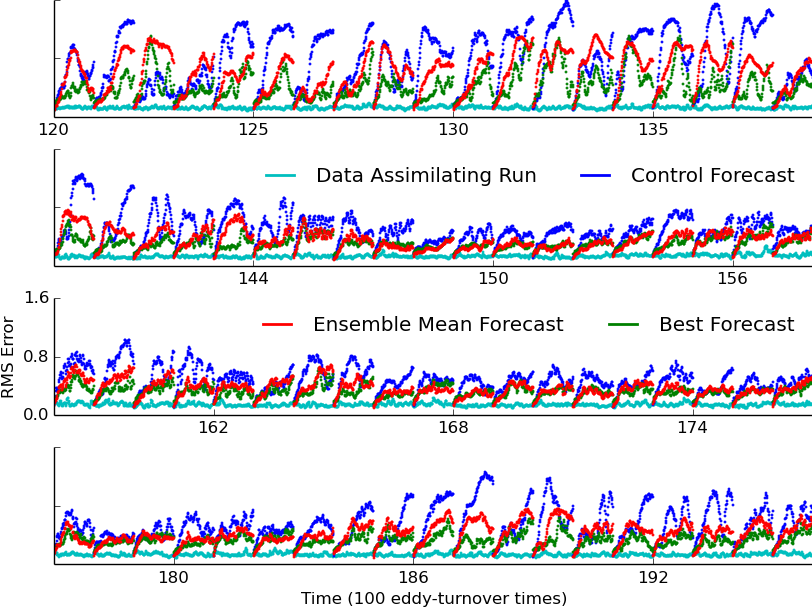}
\caption{RMS error as a function of time between times 12000 and
  20000. Error of the data assimilating run is shown in cyan, error of
  the control forecast in blue, error of the bred vector ensemble-mean
  in red and error of the best forecast member in green. When the
  system is in the zonal regime, as between times 15000 and 18000
  approximately, error is seen to be smaller than when the system is
  in the dipolar regime.
}
\label{fe}
\end{figure}

\section{Predictability in Zonal and Dipolar Regimes  }
In order to examine the predictability of the model we first generate
a truth trajectory which is simply a long run of the model over at
least one regime transition (see Fig.~\ref{k1}).  In order to mimic a
simple data assimilation scheme, the large scales ($1\le k \le 5$) of
both control and perturbed trajectories are nudged every eddy turnover
time to the true state, with the strength of nudging set so that the
error of the control forecast was comparable on average to the
standard deviation of the truth field at a lead time of about 30 eddy
turnover times (see Fig. \ref{fe}). Error is defined at the forecast
time as the root mean square of the difference between the truth and
control fields and normalized by the root mean square of the truth
field.

Stated in a different way, we are interested in predicting the true
trajectory (at a certain lead time) in a situation where
data-assimilation alone is incapable of achieving this. This is not
because data-assimilation is ineffective: the strength and other
characteristics of the periodic nudging that we use to mimic
data-assimilation measures the effectiveness of data-assimilation. To
mimic the limits of observing systems (large spatial scales and slow
temporal scales are better observable while small spatial scales and
fast temporal scales are not), we nudge only the large spatial scales
and only at every eddy turnover time. On considering the smallness of
errors in the data-assimilating run in Fig.~\ref{fe} and the left
panel of Fig.~\ref{fe-sp1}, we see that data-assimilation is
effective.  We will refer to the nudged control as the control
analysis.  A small forecast ensemble (control plus 14 perturbed
members) is constructed about the control forecast using bred
perturbation vectors (\cite{toth1997ensemble,o2008comparison}).  The
bred vectors (BVs) were obtained using the method described above and
initialized with random perturbation added to the
control. Every rescaling period, the vector difference between the
perturbed and control runs are rescaled to the initial amplitude using
the RMS-norm and added to the control run.

Figure \ref{fe} shows RMS error as a function of time between times
12000 and 20000. In this plot, error of the data assimilating run is
shown in cyan and is seen to be a small fraction (about 16\%) of the
natural variability. Error of the control forecast is shown in blue
and is seen to be dependent on whether the flow is in a zonal or a
dipolar state. In general, the forecast error is seen to be smaller
when the system is in a zonal state (approximately between times 15000
and 18500). Error of the bred vector ensemble-mean is shown in red and
is seen to be less than that of the control forecast. However, further
analysis of the ensemble behavior is necessary to assess the utility
of the ensemble prediction.

To partly address this, the error of the best forecast member is shown
in green and it is seen that at least for some of the bred vector
ensemble members, there are large reductions in forecast error.  To
further understand the improvement of forecast skill in individual
members, we examined the spatial structure of the associated bred
vector perturbations.  On comparing the evolved bred vector and the
error of the control forecast, we see a high degree of correlation
between them, suggesting that the bred vector is effectively spanning
the relevant low dimensional subspace of growing errors. This is the
reason for the improved skill of certain of the ensemble
members. It should be noted, however, that there are other
members of the ensemble that end up spanning other subspaces than
those related to errors in the control forecast. This reiterates the
necessity of further analysis of the ensemble behavior to assess the
utility of the ensemble prediction.

\section{Ensemble behavior in terms of error and spread}
\begin{figure}
\centering
\includegraphics[width=.52\textwidth]{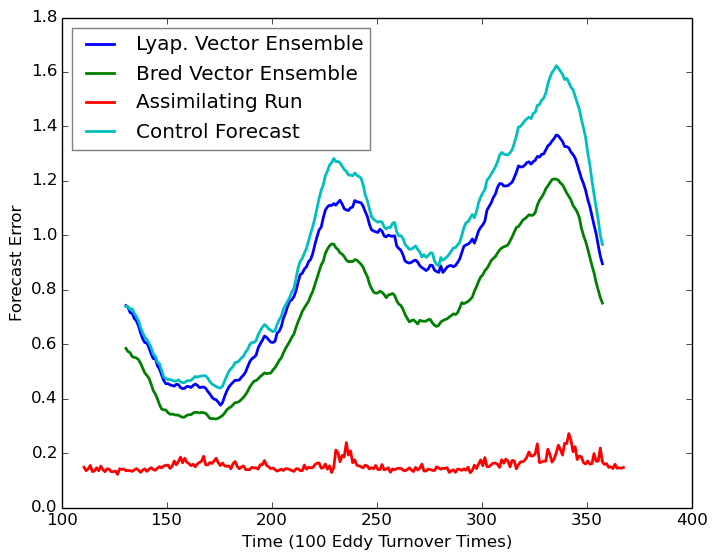}\hspace{-.9cm}
\includegraphics[width=.52\textwidth]{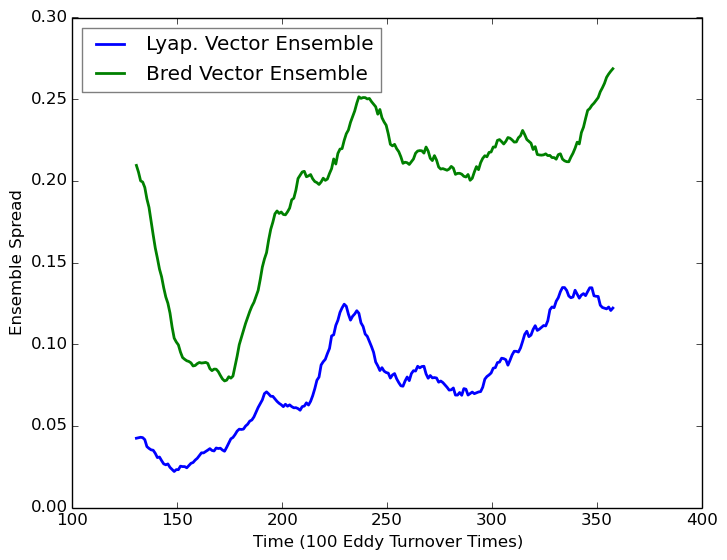}
\caption{Forecast error (left) and ensemble spread (right) averaged
  over the forecast period as a function of time.  Error of the data
  assimilating run is shown in red, error of the control forecast in
  cyan, error of the bred vector ensemble-mean in green and error of the
  Lyapunov vector ensemble-mean in blue. Also note that {\em analysis}
  error is shown for the assimilating run (even though the $y$-axis
  label is 'Forecast Error'). This is true in the other plots of
  'error' as a function of time that follow.}
\label{fe-sp1}
\end{figure}

\begin{figure}
\centering
\includegraphics[width=\textwidth]{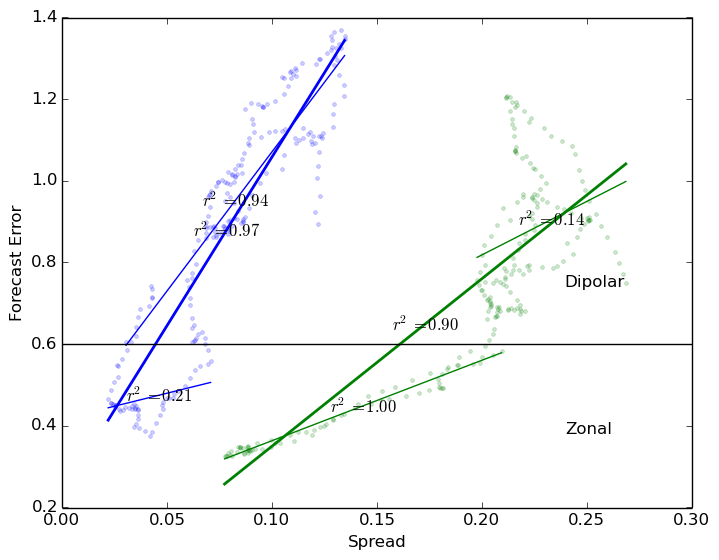}
\caption{Scatter plot of forecast error against ensemble spread. Both
  quantities are averaged
  over the forecast period. LV ensemble is shown in blue and the BV
  ensemble in green}
\label{fe-sp2}
\end{figure}

\begin{figure}
\centering
\includegraphics[width=\textwidth]{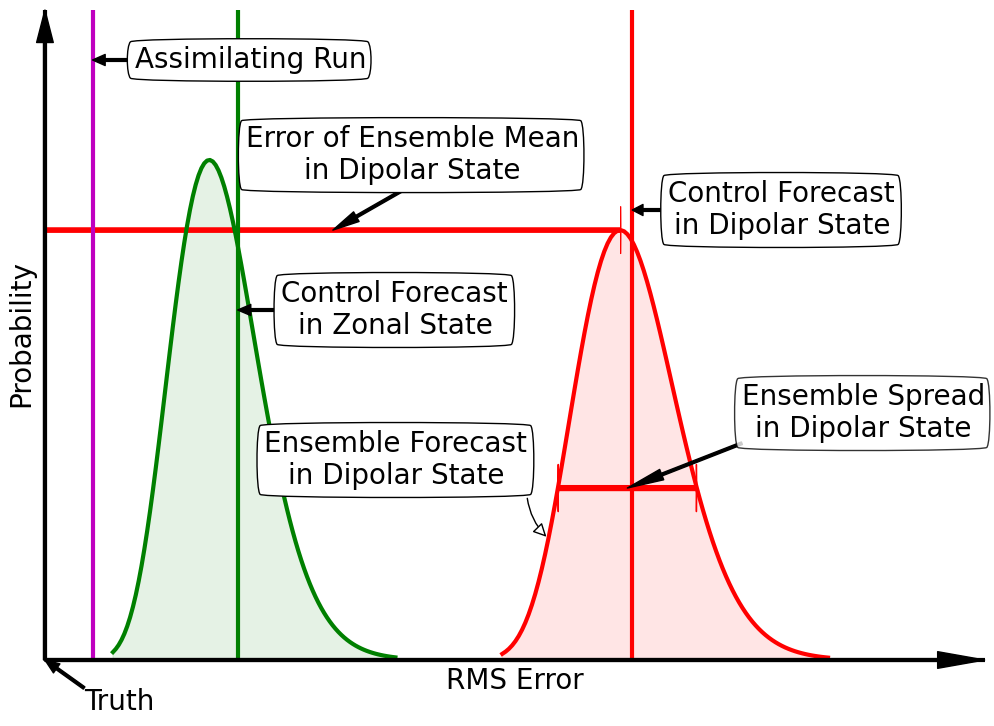}
\caption{Schematic of the error-spread relationship realized in the
  Lyapunov vector ensemble. Reduction in error in both zonal and
  dipolar regimes is small. In the dipolar regime, even though the
  reduction in error is small and the error itself large, there is
  only a marginal increase in ensemble spread over that in the zonal regime.
}
\label{SS-LV}
\end{figure}

\begin{figure}
\centering
\includegraphics[width=\textwidth]{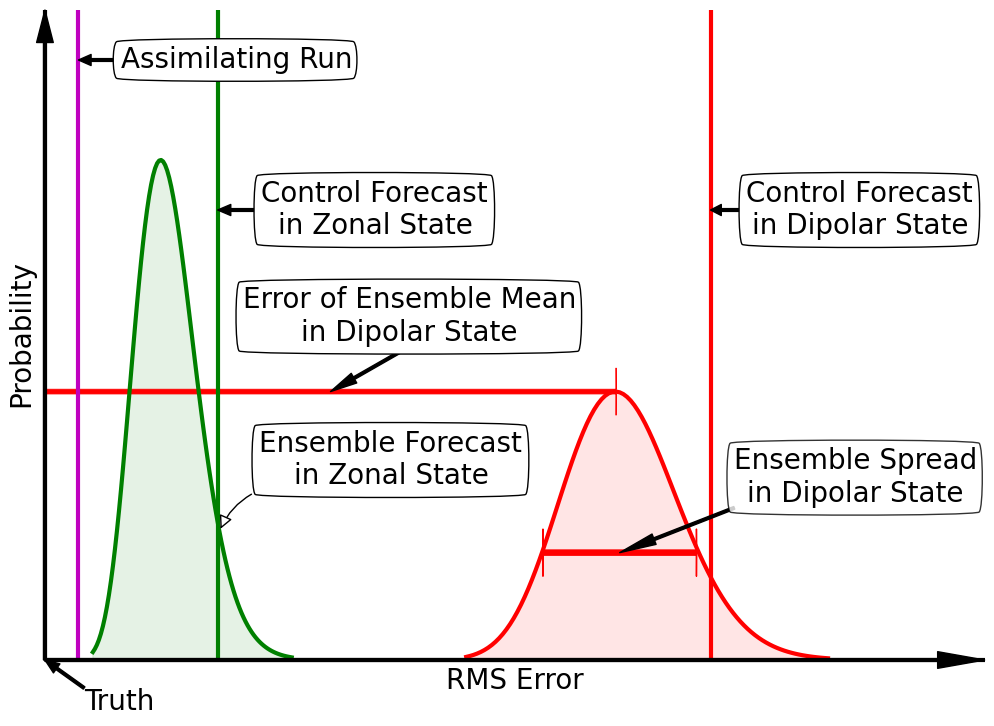}
\caption{Schematic of the error-spread relationship realized in the
  Bred vector ensemble. A larger reduction in error, as compared to
  the LV ensemble, is seen in both regimes. The larger spread of the
  ensemble members in the dipolar regime as compared to the zonal
  regime identifies lower predictability in the dipolar regime.
}
\label{SS-BV}
\end{figure}

It is traditional to consider the error-spread behavior of an ensemble
prediction system to assess its utility. In addition to this, we find
that in the present setting of a system that allows for multiple
regimes, it is helpful to interpret the error-spread relationship by
also examining individual plots of forecast error and spread as functions
of time. This is presented in  Fig.~\ref{fe-sp1}. 

In the left panel of Fig.~\ref{fe-sp1}, it is seen that the forecast
error is reduced, in both zonal and dipolar states, by a significant
fraction in the bred vector ensemble whereas this reduction is smaller
in the Lyapunov vector ensemble. The right panel of Fig.~\ref{fe-sp1}
shows spread as a function of time for the two ensembles. The larger
spread of the BV ensemble suggests greater ensemble diversity than in
the LV ensemble.  Further, when the system is in the dipolar regime,
the BV ensemble displays higher spread than when the system is in the
zonal regime. The BV ensemble is thus telling us that there is a
higher degree of uncertainty in predicting the state of the system when
the system is in the dipolar regime or likely that the predictability
of the system when in the dipolar regime is poor. The LV ensemble
corroborates this assessment of predictability in the two regimes.

Figure~\ref{fe-sp2} shows a scatter plot of forecast error as a
function of ensemble spread. Both
  quantities are averaged
  over the forecast period. LV ensemble is shown in blue and the BV
  ensemble in green
In this plot, the best-fit linear regression
lines are also shown separately for the zonal and dipolar regimes and
for the two regimes combined, and
the $r^2$ values for the linear fits are indicated, where
$$r^2 = 1 - \frac{SS_{fit}}{SS_{tot}}.$$
Here $SS_{fit}$ refers to the squared sum of the residuals with
respect to the linear fit, $SS_{tot}$ refers to the squared sum of the
residuals with respect to the mean, and $r^2$ measures the
``goodness'' of the linear fit. $r^2$ is provided to avoid
over-interpreting the fit: For example, in the zonal regime, while the
linear fit is practically useless for the LV ensemble, it is
meaningful for the BV ensemble. Finally, even when $r^2$ values are
reasonably large, systematic deviations of the scatter plot from the
linear fit is a further measure of inappropriateness of the linear fit
for the data.

The increased error and spread in the dipolar regime as compared to
the zonal regime for the BV ensemble reiterates the reduced
predictability of the dipolar regime as compared to the zonal
regime. While the differing relationship between error and spread in
the dipolar and zonal regimes is seen in the LV ensemble as well, this
difference is more muted. That is, the increased error in the dipolar
region is accompanied by only a modest increase in spread. This
suggests the possibility that the LV ensemble is less efficient as
compared to the BV ensemble.  Indeed, the behavior of the LV and BV
ensembles may be described schematically with Figs.~\ref{SS-LV} and
\ref{SS-BV}. In the LV ensemble (Fig.~\ref{SS-LV}) reduction in error
in both zonal and dipolar regimes is small. In the dipolar regime,
even though the reduction in error is small and the error itself is
large, the ensemble spread increases only marginally over that in the
zonal regime.  However, in the BV ensemble (Fig.~\ref{SS-BV}) reduction
in error in both zonal and dipolar regimes is larger than in the LV
ensemble. Further, in the dipolar regime, the larger spread of the
ensemble in the dipolar regime as compared to the zonal regime
emphasizes lower predictability in the dipolar regime. That is, the
different predictability characteristics of the zonal and dipolar
regimes are borne out better by the differing error-spread
relationships in the two regimes by the BV ensemble.

\section{Flow stability and sensitivity of BVs to rescaling amplitude
  and period}
\begin{figure}
\centering
\includegraphics[width=.49\textwidth]{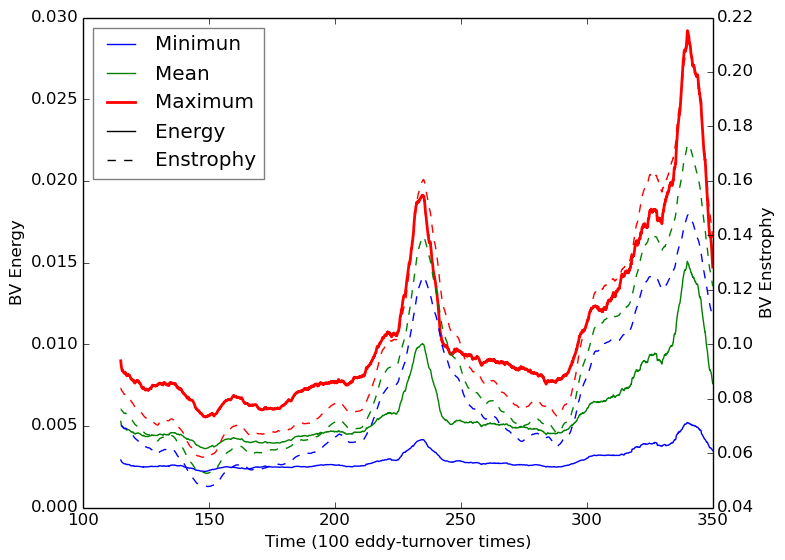}
\includegraphics[width=.49\textwidth]{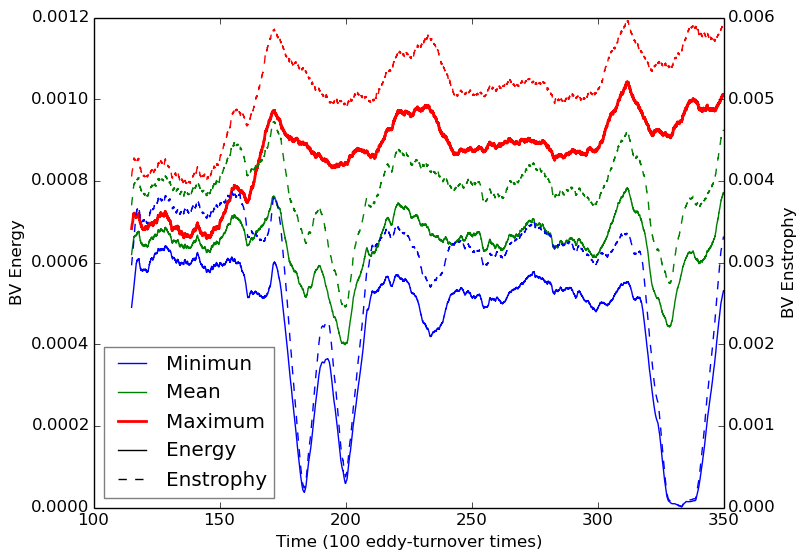}
\caption{Minimum (blue), mean (green), and maximum (red) energy
  (solid) and enstrophy (dashed) over the ensemble of BV (left) and LV
(right) perturbations. Maximum BV energy is seen to correlate best
with forecast error.}
\label{grwth}
\end{figure}

As discussed previously, predictability is associated with the
stability of the flow with respect to perturbations (errors) and their
associated growth. Figure~\ref{grwth} shows the energy and enstrophy
of the BV perturbations in the left panel and those of the LV
perturbations in the right panel. First, forecast error in
Fig.~\ref{fe-sp1} is seen to correlate better with the maximum of the
BV energy and enstrophy than with those of the LV
perturbations. Indeed, we suspect that this correlation is
causal. That is, increased forecast error when the system is in the
dipolar regime is likely due to increased instability of flow in that
regime. Further, the dominant similarity of energy and enstrophy norms
of the LV perturbations suggests that the LV perturbations are largely
concentrated at a single scale. In contrast, a variable relationship
between enstrophy and energy norms of the BV perturbations suggests
that these perturbations are expressed on a wider range of scales that
change dynamically.

\begin{figure}
\centering
\includegraphics[width=\textwidth]{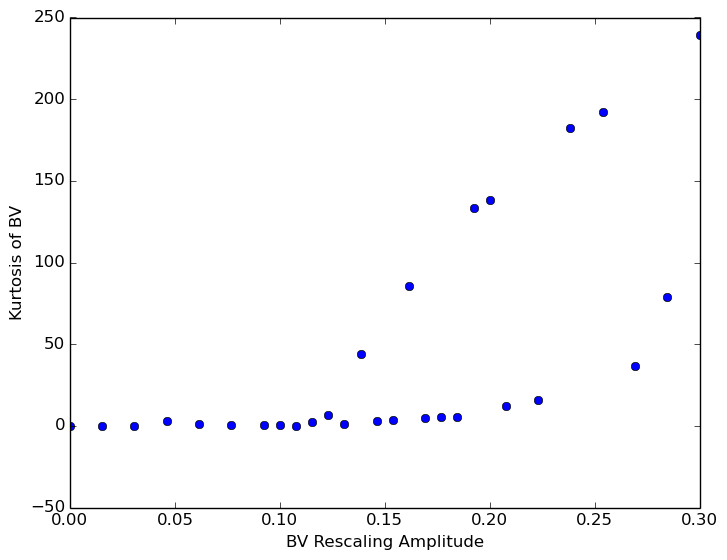}
\caption{Kurtosis of BV perturbation as a function of rescaling
  amplitude. A pitchfork bifurcation is seen to occur when the
  rescaling amplitude is about 0.12.}
\label{kurt}
\end{figure}

\begin{figure}
\centering
\includegraphics[width=\textwidth]{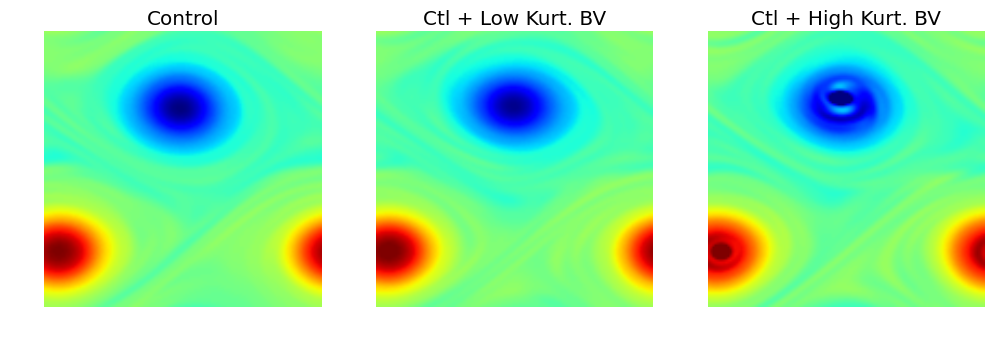}\\
\includegraphics[width=\textwidth]{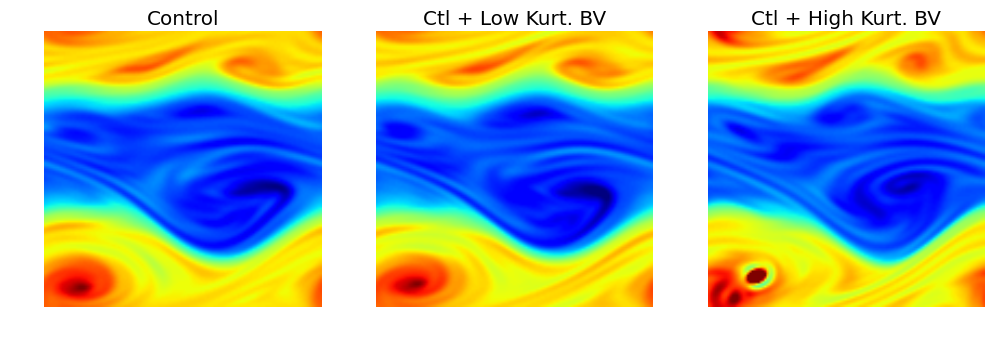}
\caption{Low and high kurtosis bred vectors in the dipolar state (top)
  and zonal state (bottom)}
\label{ctl+bv}
\end{figure}

While the expression of BVs is typically enhanced in localized regions
of large shear, it is difficult to develop an intuitive understanding
of the structure of BVs for the following reason: On examination of
various pairs of BVs, we find that not only do visually different bred
vectors lead to significantly different forecasts (not surprising,) but
that visually identical-looking bred vectors lead to significantly
different forecasts. For this reason, we consider some integrated
aspects of the BVs to further analyze them. Previously, we have
considered energy and enstrophy norms of the BV and LV
perturbations. Presently, we consider kurtosis of the perturbations.
Kurtosis is defined in the usual fashion as the deviation from a
Gaussian distribution of the fourth moment of the spatial distribution
of vorticity perturbation.  Figure~\ref{kurt} shows the kurtosis of
the bred vector perturbations for a range of rescaling amplitudes, and
a pitchfork bifurcation is seen to occur when the perturbation
amplitude is about 0.12. At this bifurcation, a branch with high
kurtosis comes into existence. Further analysis of this branch
suggests that it corresponds to a family of instability that involves
a localized distortion of the most cyclonic or anticyclonic regions of
the flow. Control vorticity fields and the sum of control field and
representative low and high kurtosis bred vectors when the system is
in the dipolar regime (top row) and zonal regime (bottom row) are
shown in Fig.~\ref{ctl+bv}.

\subsection{Dependence of error and spread on perturbation rescaling period}
\begin{figure}
\centering
\includegraphics[width=.52\textwidth]{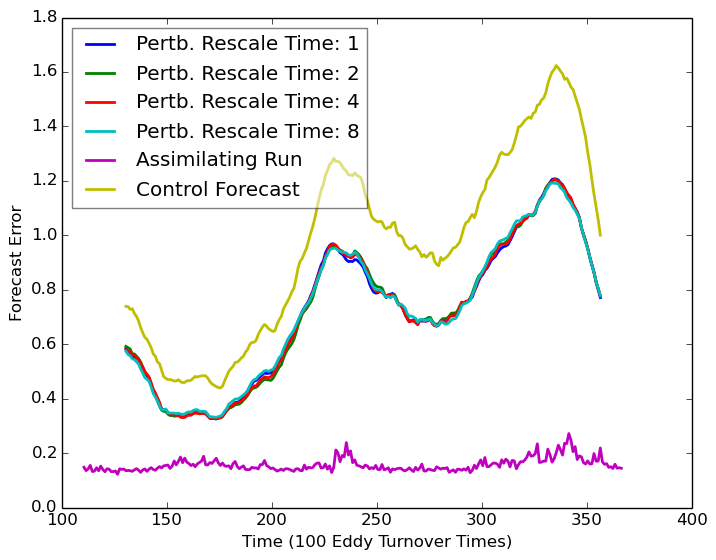}\hspace{-.9cm}
\includegraphics[width=.52\textwidth]{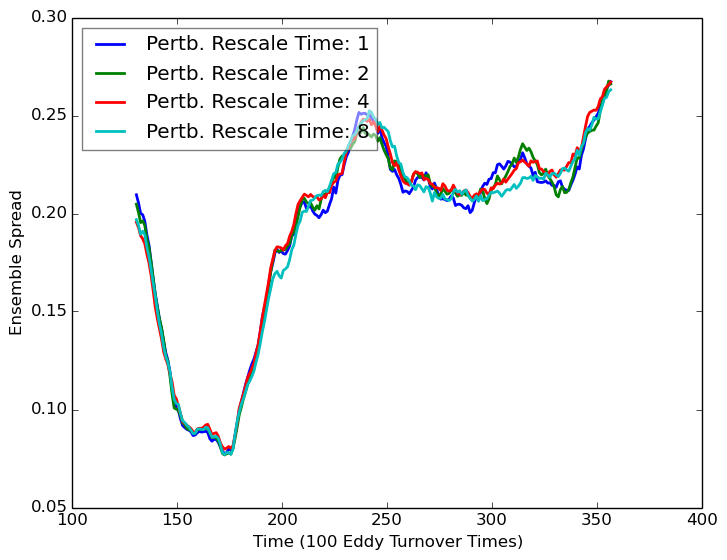}
\caption{Forecast error (left) and ensemble spread (right) averaged
  over the forecast period as a function of time. In this series of
  four experiments, the amplitude of the perturbation vectors are
  rescaled at 1, 2, 4, and 8 eddy turnover times and the amplitude to
  which the perturbations are rescaled is large (0.2 times the
  initial base enstrophy). In this case of finite amplitude bred
  vector perturbations, the dependence of forecast error and ensemble
  spread on rescaling time is seen to be very weak.
}
\label{fe-sp-p2}
\end{figure}

\begin{figure}
\centering
\includegraphics[width=.52\textwidth]{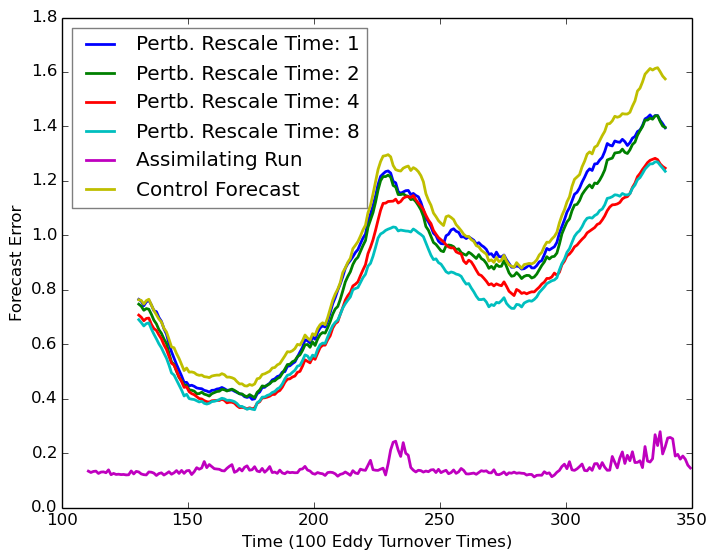}\hspace{-.9cm}
\includegraphics[width=.52\textwidth]{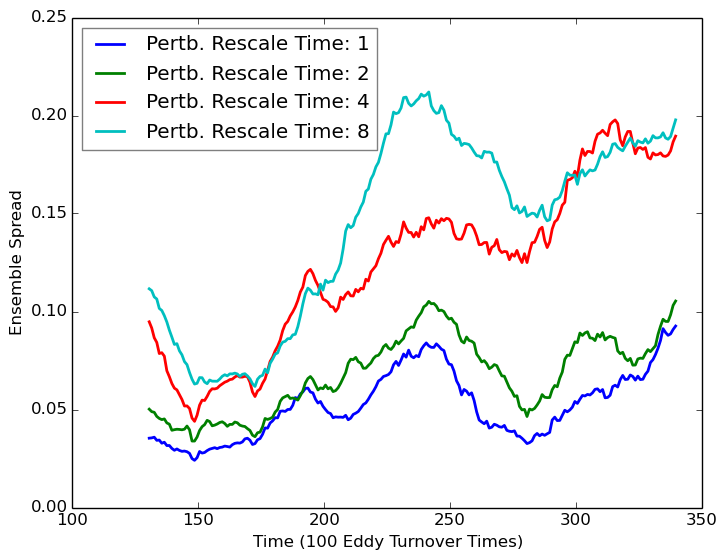}
\caption{This series of four experiments are similar to those in
  fig.~\ref{fe-sp-p2}: The only difference is that the amplitude to
  which the perturbations are rescaled is small (10$^{-5}$ times the
  initial base enstrophy). In this case, the dependence on rescaling
  time is strong: The decreased error and increased spread in the two
  cases where the rescaling times are 4 and 8 eddy turnover times is
  likely due to the perturbations having grown large and a
  manifestation of the nonlinear filtering effect of finite-amplitude
  evolved perturbations.}
\label{fe-sp-1em5}
\end{figure}

In Fig.~\ref{kurt}, we saw how different kinds of instabilities are
captured depending on the rescaling amplitude of the BVs. Likewise, it
is possible that different growing modes of the system are captured
depending on the rescaling period for the BVs. In a shear-dominated
flow as we presently consider, this could be for example due to
reasons of non-normality of the linear operator. Or such dependence
could be expected because of nonlinear saturation kinds of
effects. For this reason, a series of experiments were conducted
wherein the rescaling period was varied between 1 and 8 large-scale
eddy turnover times.  Figure.~\ref{fe-sp-p2} shows the forecast error
(left) and ensemble spread (right) averaged over the forecast period
as a function of time for the case where the rescaling amplitude is
large . In this series of four experiments, the amplitude of the
perturbation vectors are rescaled at 1, 2, 4, and 8 eddy turnover
times and the amplitude to which the perturbations are rescaled is
large (0.2 times the initial base enstrophy). In this case of finite
amplitude bred vector perturbations, the dependence of forecast error
and ensemble spread on rescaling time is seen to be very weak.  On the
other hand, when the same experiments were repeated for the case when
the rescaling amplitude is smaller, (10$^{-5}$ times the initial base
enstrophy), a stronger dependence of error and spread on rescaling
time is seen (Fig.~\ref{fe-sp-1em5}. The decreased error and increased
spread in the two cases where the rescaling times are 4 and 8 eddy
turnover times is likely due to the perturbations having grown large
and a manifestation of the nonlinear filtering effect of
finite-amplitude evolved perturbations.

\subsection{Error and spread of random rescaling BV perturbations}
\begin{figure}
\centering
\includegraphics[width=.52\textwidth]{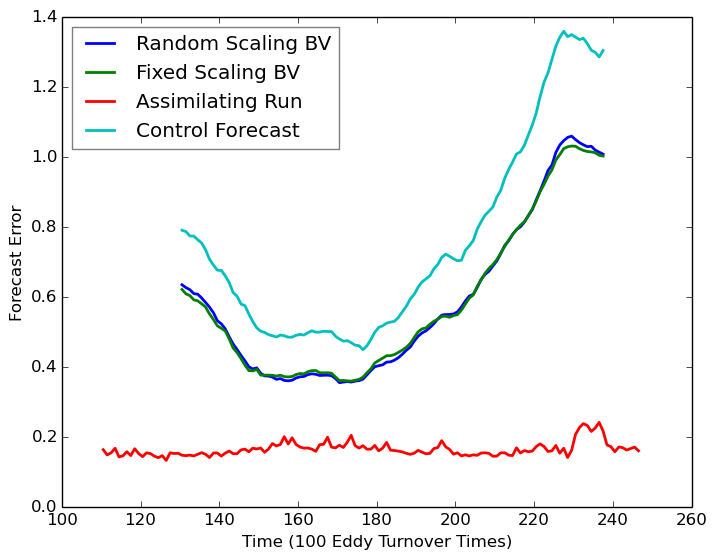}\hspace{-.9cm}
\includegraphics[width=.52\textwidth]{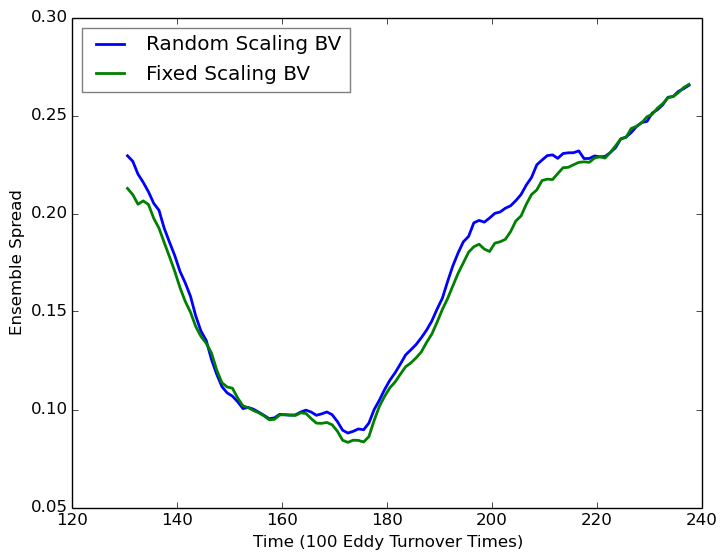}\\
\includegraphics[width=.52\textwidth]{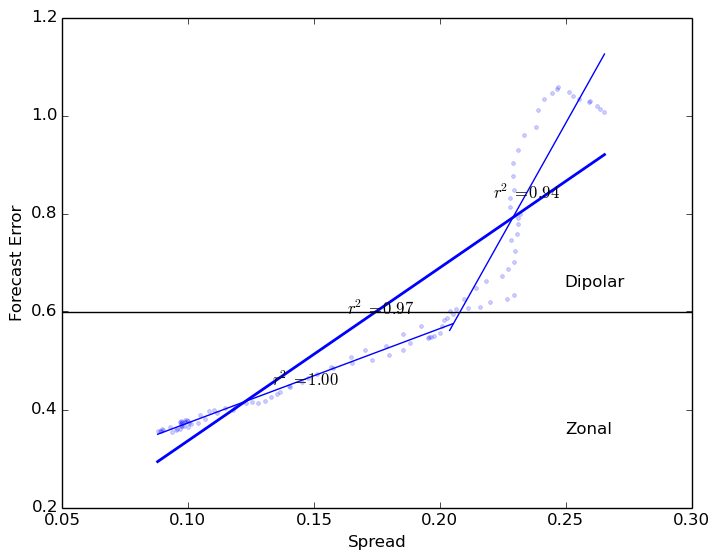}
\caption{Random rescaling versus fixed rescaling BV ensembles. When
  both rescaling amplitude and rescaling period are drawn from uniform
  distributions (see text for ranges), the ensemble behavior is
  similar to that of an ensemble with fixed (finite) rescaling
  amplitude and fixed (long) rescaling period.}
\label{fe-sp-rr}
\end{figure}

We finally consider an ensemble of BV perturbations in which the
rescaling amplitude and rescaling period are both chosen randomly;
rescaling amplitude was chosen based on an enstrophy norm from an
uniform distribution between $10^{-5}$ and $0.2$ times the initial
base enstrophy and rescaling period was chosen randomly from an
uniform distribution between 1 and 8 eddy turnover times. It is seen
in Fig.~\ref{fe-sp-rr} that the behavior of this ensemble is not
significantly different from a BV ensemble in which the rescaling
amplitude and rescaling period are both held constant (rescaling
amplitude of 0.2 times base enstrophy and a rescaling period of 8
turnover times). Thus given the robustness of error and spread
behavior for finite amplitude perturbations, it seems that guesswork
may be largely removed from the use of BVs by choosing such a random
rescaling strategy, as long as the range of rescaling parameter values
are chosen reasonably. However, this finding needs to be verified
before being used in other systems.

\section{The role of small scales}
It was anticipated that the higher order non-Gaussian nature of the
distribution of vorticity that at least some of the larger amplitude
bred vectors exhibited may be of importance in identifying small scale
coherent error structures and therefore play a dynamical role in
predictions.  In order to examine this possibility, we conducted
experiments in which nudging was also considered at small scales.
That is, whereas in the default setting, assimilation was performed
only at the large scales, a second setting was also considered wherein
assimilation was performed at all scales. The hypothesis here being
that when assimilation is performed at small scales, the coherence of
dynamically important structures (that possess small scale features)
in the BV perturbations is broken, thus reducing the effectiveness of
BV perturbations in representing such structures.  

\begin{figure}
\centering
\includegraphics[width=.52\textwidth]{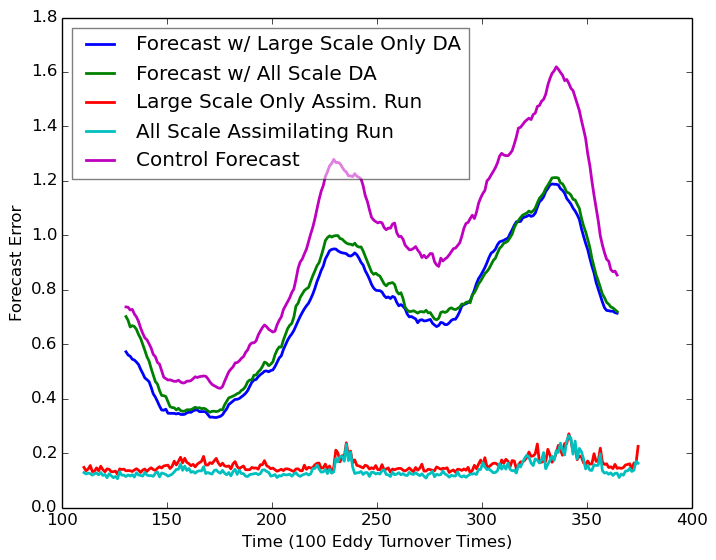}\hspace{-.9cm}
\includegraphics[width=.52\textwidth]{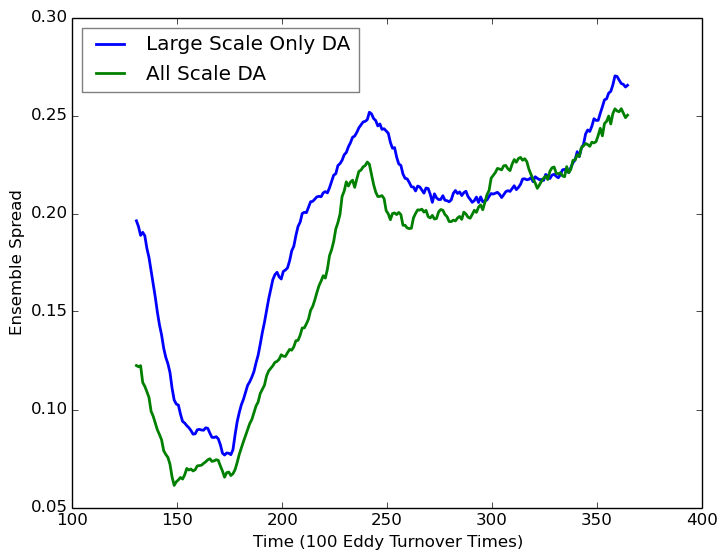}\\
\caption{When data assimilation is applied at all scales (green), it
  is seen that the forecast error increases slightly whereas the
  ensemble spread mostly decreases as compared to when data
  assimilation is applied only to the large scales (blue). This
  points to the dynamical importance of small scales to improving
  predictions.}
\label{da}
\end{figure}

Figure.~\ref{da} shows the forecast error (left) and ensemble spread
(right) when assimilation is performed at all the scales in the system
(green), rather than at  only the large scales of the system (blue).
Firstly, among the two assimilating runs, it is seen that when
assimilation is also performed at small scales, analysis error is
reduced (in comparison to the default case wherein assimilation is
only performed at large scales). However, error in the two control
forecasts initialized with the two differently assimilating (large
scale only versus all scale) is far more similar. This suggests
ineffectiveness of performing assimilations at smaller scales. (Such
ineffectiveness is likely related to the dynamics of backscatter---or
how the small scales affect the large scales; e.g., see
\cite{nadiga2008orientation}). That is, even though error at small
scales is held down by data assimilation at those scales, the
incremental reduction in analysis error over the (default)
assimilation setup wherein assimilations are only performed at large
scales, does not translate into improved predictions.

In the ensemble forecast context, it actually so happens that
performing data assimilation at the smaller scales has a {\em
  deleterious} effect. This can be seen in the slight increase in
ensemble mean forecast error at most times when assimilation is also
performed at small scales. Finally, the increased ensemble-mean
forecast error is accompanied by reduced ensemble spread. This
suggests that that in the case when smaller scales are assimilated,
the diversity of the BVs is actually reduced. Such combined
behavior of error and spread highlights the importance of considering
the role of instabilities at small scales in flow predictions.  When
restoring is present at the small scales, instabilities at these
scales or the coherent development of structures that include these
scales are inhibited in BVs. This either precludes or partially
thwarts the ability of BVs to highlight incipient dynamically
important regions and evolve the associated structures to compensate
co-located control forecast errors.

\begin{figure}
  \centering
\includegraphics[width=\columnwidth]{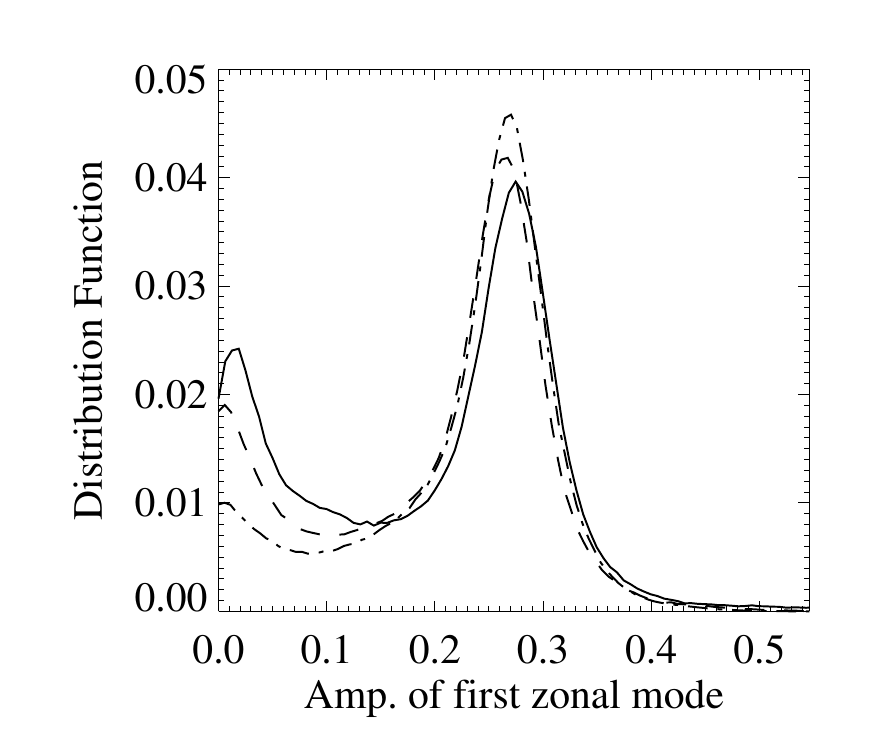}
\caption{Distribution function of amplitude of energy in the first
  zonal mode.  When the proportion of forcing at small scales is
  increased progressively from solid to dashed to dot-dashed, the
  dipolar state is increasingly preferred and transitions between the
  zonal and dipolar states become less frequent.}
\label{pdf}
\end{figure}

\section{Discussion and conclusion}
We considered a model based on the barotropic vorticity equation and
in which both dissipation and stochastic forcing are weak compared to
nonlinearity. (Note that weakly dissipative regimes are generally
inaccessible to state of the art general circulation models because of
resolution requirements.) This system exhibits the phenomenon of
low-frequency regime transitions.  A dynamical explanation of
low-frequency regime transitions in a system is as follows: if the
system supports multiple equilibria, then noise can induce transitions
between such equilibria through a form of stochastic resonance. Zonal
flows and dipoles are well recognized as statistical mechanical
equilibria of the system we consider. Then, specifically, at any given
time, the system is in a basin of attraction of a particular attractor
(regime). Cumulative effect of stochastic forcing, however, pushes the
system over the attractor boundary so that the system finds itself in
the basin of attraction of a different attractor (different
regime). Dynamical processes then lead to a relaxation of the system
onto the different attractor (regime) and the process continues.

We were interested in the predictability of flow in the different
regimes and in the role small scales play in such predictions. We were
also interested in the role small scales play in low-frequency regime
transitions. To this end, we conducted prediction experiments that use
a simple data-assimilation scheme. Indeed the process of data
assimilation is emulated by nudging the large scales of the system to
truth on the large eddy turnover timescale. While the estimate of the
state of the system using such a scheme was good, predictions based on
that estimate alone (control forecast) were deficient.  We then
examined the use of bred vector perturbations and Lyapunov vector
perturbations in improving the control forecast. We find that
ensembles that use finite amplitude BV perturbations are more
effective at reducing forecast error and are more reliable as compared
to ensembles that use LV perturbations. This is consistent with
previous understanding and explanations of predictability in
multiscale systems that emphasize the importance of the growth of
finite amplitude perturbations as opposed to purely linear
instabilities. Indeed, the advantages of BV perturbations that we
demonstrate in this article are automatically realized when data
assimilation is based on a genuine dynamically evolving ensemble
\citep[e.g., see][and references therein]{nadiga2013ensemble}.

Focusing further attention on finite amplitude BV perturbations, we
found that forecast errors were smaller than that of the control
forecast for some of the members. For such members, the error
reduction was seen to be achieved by a process wherein dynamically
important unstable subspaces identified by BV perturbations evolve
into structures that compensate regions of control forecast error.

Next, we found that the system has lesser predictability when it is in
the dipolar regime as compared to the zonal regime. Indeed the peaks
of vorticity that are characteristic of the dipolar regime make the
flow in that regime more unstable and constitutes the primary reason
for lesser predictability of the system when in that regime. This
instability is however unlikely to be the usual asymptotic linear
instability alone. Our experiments suggest that instability of finite
amplitude perturbations play an equally important role. In ensemble
prediction experiments that use finite amplitude BV perturbations and
those that use LV perturbations, forecast errors of ensemble
predictions are seen to correlate far better with measures of
instability as provided by the finite amplitude BV
perturbations. Indeed, by systematically varying the amplitude of the
perturbations, evidence for a new branch of instability at a finite
amplitude was presented. Nevertheless, the horizon of predictability
in either of the regimes was short compared to characteristic time
scales associated with processes that lead to regime transitions, thus
precluding the possibility of predicting such transitions.

The ensemble predictions were also analyzed for their utility by
examining forecast error and ensemble spread and their relationship in
the two regimes. The error-spread relationship was seen to be
different in the two regimes with a much stronger linear relationship
with a positive correlation being realized when the system was in the
zonal regime.  The most uncertain aspect of prediction when the system
is in the dipolar state is the position of peak cyclonic and
anti-cyclonic vorticities and the ensemble-mean averages over
this uncertainty. Consequently, while there was a reduction in error
of the ensemble-mean over that of the control forecast, it was, in a
sense, not very useful or skillful. This explains the weaker
relationship between error and spread in the dipolar regime. Further,
smaller error and better error-spread relationship was realized with
the finite amplitude BV ensemble than with the LV ensemble.

A further series of experiments demonstrated that as long as the
amplitude of the perturbations where large enough (say to capture the
new branch of instability that was identified), the behavior of the
ensemble was robust to rescaling period and amplitude. Indeed the
ensemble behavior with a random rescaling strategy was statistically
no different from that with a fixed rescaling strategy. Next, that the
small scales play a significant role in predictions was established by
conducting experiments in which the small scales were either included
or not included in the simple data assimilation procedure. In
experiments that nudged the small scales as well, the ability of BV
perturbations to carry information about dynamically important
structures was likely compromised leading to larger forecast errors and
smaller ensemble spread. This led us to form a hypothesis about the
role of small scales in initiating low-frequency regime transitions:
If the small scales play a role in initiating low-frequency
transitions, then extending stochastic forcing to the smaller scales
should suppress transitions. Figure~ \ref{pdf} shows that in experiments where
forcing was confined to small scales or where a larger proportion of
the forcing was at small scales, no transitions were observed. In
experiments where transitions were observed, the transitions became
infrequent as the proportion of forcing at small scales (still small)
increased. These experiments therefore verify the importance of small
scales and the process of backscatter in initiating low-frequency
regime transitions in the model.

The relevance of this study to the ocean-atmosphere system is twofold:
a) It is well recognized that the timescales for low frequency
variability in the climate system are largely set by the ocean with
the atmospheric state acting as a source of stochasticity. However,
the atmospheric forcing can be correlated at large spatial scales
(atmospheric teleconnections). The role of such stochastic forcing in
determining the ocean state is only beginning to be considered
\cite[e.g., see][]{o2013decadal}.  This study may
be seen as preliminary work towards such a characterization. Thus if (as
is likely) ocean circulation is predisposed to supporting multiple
equilibria, then weak stochastic forcing by the atmosphere has the
potential to induce transitions in the ocean state by a form of
stochastic forcing.  b) We find that coherent evolution of small scale
structures likely plays a role in initiating low-frequency transitions
between states with very different large scale structure and that
accounting for the presence of such dynamically important small scale
structures is important for predictions. An implication of this
finding for models that do not explicitly resolve such small scales,
but rather parametrize them in a generic fashion \citep[e.g., see][and
references therein]{frederiksen2008entropy}, is a likely reduction in
its predictive capability.

\section{acknowledgments}
TJO is supported by an Australian Research Council Research
Fellowship.


\bibliographystyle{cambridgeauthordate}
\bibliography{cup}
\label{refs}

\end{document}